\documentstyle[preprint,revtex]{aps}

\begin{document}
\thispagestyle{empty}

% TITLE PAGE

\begin{center}
\hfill{IP-ASTP-18-94}\\
\hfill{July 1994}

\vspace{1 cm}

\begin{title}
Quantum Mechanics and Linearized Gravitational Waves
\end{title}
\vspace{1 cm}

\author{A.~D.~Speliotopoulos}
\vspace{0.5 cm}

\begin{instit}
Institute of Physics, Academia Sinica\\ Taipei, Taiwan 115, R.O.C.
\end{instit}
\end{center}
\vspace{0.5 cm}

\begin{abstract}
\vspace{1 cm}

The interaction of classical gravitational waves (GW) with matter is
studied within a quantum mechanical framework. The classical
equations of motion in the long wave-length limit is quantized and a
Schroedinger equation for the interaction of GW with matter is
proposed. Due to its quadrapole nature, the GW interacts with matter
by producing squeezed quantum states. The resultant hamiltonian is
quite different from one would expect from general principles,
however. The interaction of GW with the free particle, the harmonic
oscillator and the hydrogen atom is then studied using this
hamiltonian.

\end{abstract}
\newpage

\noindent{\bf \S 1. Introduction.}

When the two words ``quantum'' and ``gravity'' appear together in the
same sentence, they usually appear as ``quantum gravity'', a term which
immediately brings to mind energies and length scales at the level of
the Planck scale. With the advent of LIGO, and the prospect of the
direct detection of gravitational waves (GW) $\cite{Brag}$,
$\cite{Thorne}$, however, the interplay of classical and quantum
gravity at a {\it quantum mechanical\/} level begins to take on added
importance. Indeed, from the classical analysis the effect of GW on
matter is expected to be so small that one begins to expect that a
quantum mechanical treatment of this interaction is needed
$\cite{Caves}$, $\cite{Weber}$. It is, moreover, at the quantum
mechanical level that experimental evidence for the interplay between
gravity and quantum mechanics will first appear, rather
than at the level of full quantum gravity.

Surprisingly, the interaction of GW and matter at a quantum
mechanical level has not been systematically explored. We know of two
approaches currently in the literature. The first is by DeWitt
$\cite{DeWitt}$ and is concerned solely with the general formulation
of quantum mechanics on a curved, background space(time). The second is
by Weber $\cite{Weber}$ and, as he was concerned with the response of
GW detectors to incident GW's, deals exclusively with the interaction
of GW with matter. His approach involved the direct quantization of
the {\it linearized\/} classical equations of motion for a test
particle interacting with a GW propagating on a flat
background. Surprisingly, these two approaches are {\it not\/}
equivalent to one another. One cannot, for example, obtain Weber's
result, or the results of this paper, from DeWitt's general formalism
by taking the linearized gravity limit. Fundamentally, we shall find
that this is due to the different approaches taken by Weber and
DeWitt. Weber begins with essentially the geodesic deviation equation
which he then quantizes while DeWitt does a straightforward
generalization of the quantization proceedure in flat spacetime to
curved spacetimes.

In this paper we shall present a systematic study of the interaction
of GW with matter at a quantum mechanical level in the long
wave-length and small velocity limit. Like Weber, we shall start with
the classical equations of motion, but unlike his approach we shall
not first linearize the equations of motion. We find
that the hamiltonian derived by Weber is valid only under very
restrictive  circumstances and, in particular, is not valid if the
test particle is charged. In fact, due to the quadrapole
nature of the GW (instead of the dipole nature of the
electro-magnetic (em) field), the net effect of an incident GW on any
quantum mechanical system is to produce a squeezed quantum state.
This aspect of the interaction could not be seen within the
linearized approach of Weber. Another aspect is that the
GW is found to couple with test particle in a way that is quite
similar to the minimally coupled em-field. While on the one hand this
is what we would expect if the analogy between GW and em-waves is to
hold, on the other hand it is quite different than what one
expects from general principles. Spin-$0$ scalar particles like the
ones we shall be concerned with in this paper are not expected to
couple to the GW in this way.

The rest of this paper is organized in the following manner. In
$\bf\S 2$ we shall quantize the classical theory using the usual
canonical quantization proceedure. Then, in the succeeding sections
$\bf \S 3$ to $\bf \S 5$ we shall use this hamiltonian to study the
interaction of the GW with the free particle, the two dimensional
harmonic oscillator, and the hydrogen atom. Concluding remarks and
comparisons between DeWitt's, Weber's and our results can then be
found in $\bf\S 6$.

\noindent{\bf \S 2. Quantum Mechanics.}

The equations of linearized gravity on a Minkowski background are
well known. From $\cite{MTW}$,
\begin{equation}
\Gamma^\mu_{\alpha\beta} = \frac{1}{2}\eta^{\mu\nu}\left(
			    \partial_\alpha h_{\beta\nu} +
			    \partial_\beta h_{\alpha\nu} -
			    \partial_\nu h_{\alpha\beta}
			    \right)\>,
\label{e1}
\end{equation}
where $h_{\mu\nu}$ is the perturbation off $\eta_{\mu\nu}$, the flat
Minkowski metric and
\begin{equation}
R_{\alpha\beta\mu\nu} = \frac{1}{2}\left(
			\partial_\mu\partial_\beta h_{\alpha\nu}
			+
			\partial_\alpha\partial_\nu h_{\beta\mu}
			-
			\partial_\alpha\partial_\mu h_{\beta\nu}
			-
			\partial_\beta\partial_\nu h_{\alpha\mu}
			\right).
\label{e2}
\end{equation}
We shall only be concerned with GW's propagating on a flat background
and its effect on scalar, spin-$0$ particles. By choosing the
transverse and traceless gauge for the GW,
\begin{equation}
h_{0\mu} = 0\>,\qquad \nabla^\mu h_{\mu\nu} =0\>,\qquad h_\mu^\mu=0\>,
\label{e3}
\end{equation}
all the gauge freedom has been removed and the GW
contains only its two physical polarizations. The first polarization
state $h_{11} = -h_{22}$, $h_{33}=h_{12}=h_{21}=0$ is usually
called the $+$-polarization while the second $h_{12} = h_{21}$, $h_{jj} =
0$ (no sum) is usually called the $\times$-polarization. In this
gauge the only non-zero componants of the curvature tensor are
\begin{equation}
R^j_{0,k0} = - \frac{\partial \Gamma^j_{0k}}{\partial t} =
-\frac{1}{2}\frac{\partial^2h_{jk}}{\partial t^2}\>.
\label{e4}
\end{equation}
As usual, greek indices shall run from $0-4$ while latin indices run
from $1-3$ and we are following the sign convention in $\cite{MTW}$.

Classically, the response of a scalar test particle
to the passage of a gravitational wave is given by the geodesic
deviation equation. The overall effect of the GW is to introduce an
effective tidal force to the equations of motion,
\begin{equation}
m\frac{d^2 x^j}{dt^2} = - mR^j_{0,k0}x^k + F^j \>,
\label{e5}
\end{equation}
where $x^j$ is the location of the test particle, $m$ is its mass and
$F_j$ represents all other forces acting on the particle
(see $\cite{MTW}$ for a complete derivation and explanation of the
above). In the derivation of eq.~$(\ref{e5})$, certain important
approximations were made. First, it was assumed that the test
particle is slowly moving and any velocity terms in the geodesic
deviation equation may be neglected. Second, the long wave-length
limit was taken, meaning that the reduced wave-length $\lambda/2\pi$
of the GW was assumed to be very much larger than the range of motion
of the test particle. The GW can then be treated as a function of
time only.

Eq.~$(\ref{e5})$, the classical equations of motion for
the test particle, is our starting starting point. As we shall see,
there are {\it two\/} different ways of quantizing the classical
theory which are unitarily equivalent to one another only under very
restrictive circumstances. We shall start with the most
straightforward quantization procedure. For convenience, we shall
assume that any external forces are time and velocity independent and
can be represented as a potential $V$. Moreover, we shall always
treat the GW as an external, {\it classical\/} field.

The classical lagrangian for eq.~$(\ref{e5})$ is
\begin{equation}
L_{NR} = \frac{1}{2} m\dot x_j^2 - \frac{m}{2}R^j_{0,k0}x^jx^k -V\>,
\label{e6}
\end{equation}
where the dot denotes derivative with respect to time. Then choosing
$x^j$ as our generalized coordinate, the canonical momentum is $p_j =
mx^j$, the mechanical momentum, and the hamiltonian
is
\begin{equation}
H_W = \frac{p_j^2}{2m} + \frac{m}{2}R^j_{0,k0}x^k+V\>.
\label{e7}
\end{equation}
We may then ``quantize'' this hamiltonian in the usual manner by
replacing $x^j$ and $p_j$ with operators which satisfy the canonical
quantization condition: $[x^j,p_k]=i\hbar\delta^j_k$. It is
essentially a linearized version of eq.~$(\ref{e7})$ which is used by
Weber and others $\cite{Gris}$ in studying the quantum mechanical
properties of GW detectors.

The second way to proceed is to note that the lagrangian
eq.~$(\ref{e6})$ is equivalent to
\begin{equation}
L = \frac{1}{2} m\dot x^2 - m\Gamma^j_{0k}\dot x_jx^k -V\>,
\label{e8}
\end{equation}
up to an integration by parts. If we once again choose $x^j$ as our
generalized coordinate, we now find that the canonical momentum is
$P_j = m\dot x_j - m \Gamma^j_{0k} x^k$ while the hamitonian becomes
\begin{equation}
H = \frac{1}{2m}\left(P_j + m\Gamma^j_{0k}x^k\right)^2 +V\>.
\label{e9}
\end{equation}
Canonical quantization then gives: $[x^j, P_k] = i\hbar \delta^j_k$.
The Schr\"oedinger equation is then
\begin{equation}
-i\hbar\frac{\partial \psi}{\partial t} = H \psi\>,
\label{e9a}
\end{equation}
with the expectation value of any operator $\cal O$ defined as
\begin{equation}
\langle\psi\vert{\cal O}\vert\psi\rangle =\int\overline\psi {\cal
O}\psi d^3x\>.
\label{e9b}
\end{equation}
A priori, we might have expected the presence of $\sqrt{g^{(3)}}$ in
the integral eq.~$(\ref{e9b})$ where $g^{(3)}$ is the determinant of the
three dimensional metric for a hypersurface. Since, however, we are
dealing with {\it linearized\/} gravity, $g^{(3)}_{jk} = \delta_{jk}
+h_{jk}$ and $g^{(3)} \approx 1+h_{jk}^2/2$. To lowest order,
$g^{(3)}\approx1$.

Since $\Gamma^j_{0k}$ depends only on time, the quantization
condition $[x^j, P_k] = i\hbar \delta^j_k$ implies that $[x^j, m\dot
x_k] = i\hbar \delta^j_k$ also. The two hamiltonians
eq.~$(\ref{e7})$ and eq.~$(\ref{e9})$ are related to one another by a
simple canonical transformation. Thus there is a {\it time dependent\/}
unitary transformation which maps one to the other. On this level
one can just as well use either eq.~$(\ref{e7})$ or eq.~$(\ref{e9})$
to analyze the response of the test particle to a GW. How the
system is viewed physically, on the other hand, differs tremendously
from one hamiltonian to the other. Notice the striking similarity
between eq.~$(\ref{e9})$ and the hamiltonian for a charged particle
minimally coupled to the em field. Like the em-field, the GW
is expected to carry momentum, which is recognized by
eq.~$(\ref{e9})$, and like the vector potential for the em-field,
eq.~$(\ref{e9})$ implies that the connection $\Gamma^j_{0k}$
plays a fundamental role at the quantum mechanical. One should keep
in mind, however, that we are dealing with a {\it scalar\/} test
particle and for a scalar particle one does not expect the connection
to appear in the hamiltonian in this manner. Instead, at the most one
would expect a direct coupling to the metric of the spacetime, as
was obtained by DeWitt.

Let us emphasize that this simple relationship between the two hamiltonians
holds only under very restrictive conditions. For example, if the
test particle is also charged, then under minimal coupling
eq.~$(\ref{e9})$ becomes
\begin{equation}
H = \frac{1}{2m}\left(P_j -\frac{q}{c} A_j+
m\Gamma^j_{0k}x^k\right)^2 +V\>,
\label{e10}
\end{equation}
in the non-relativistic limit while
\begin{equation}
H_W = \frac{1}{2m}\left(p_j -\frac{q}{c} A_j\right)^2 +
\frac{1}{2}R^j_{0,k0}x^k+V\>,
\label{e11}
\end{equation}
where $q$ is the charge of the particle and $A_j$ is the vector
potential. Expanding out the quadratic momentum term, we find that
$H$ contains a direct interaction term between the em-field and the
GW while $H_W$ does not. Moreover, due to the presence of the
em-field the canonical momentum $P_j$ and the mechanical momentum
$m\dot x_j$ are no longer related to one another through a canonical
transformation. The two hamiltonians will, in general, describe
different physics.

Actually, this type of equivalence up to a canonical transformation
between two hamiltonians also occurs when an em-field interacts with
a charged particle. If an em-field in the
long wave-length limit interacts with a slowly moving charged
particle, then if we neglect any velocity terms in the interaction the
hamiltonian can be approximated by
\begin{equation}
H_{em} = \frac{p_j^2}{2m} + q E_j(t) x_j(t)\>.
\end{equation}
where $E_j$ is the electric field and is treated as a classical field.
Because of the long wave-length approximation, is a function
of time only. The charged particle actually couples minimally to the
em-field, of course, but in this limit the two hamiltonians are
related to one another by unitary transformation.

We should also mention that both eqs.~$(\ref{e7})$ and $(\ref{e9})$
hold only in the long wave-length limit. Neither one is expected to
hold in the general case where $\Gamma_{0k}^j$ is also position
dependent. In fact, if we arbitrarily allow $R^j_{0,k0}$ to be
position dependent, then the equations of motion calculated from
eq.~$(\ref{e7})$ will contain terms which depend on the derivative of
$R^j_{0,k0}$ which are not present in the geodesic deviation
equation. Indeed, we see that since the geodesic deviation equation
depends on the curvature, the hamiltonian calculated from it can only
depend on the connection and should therefore have a form
which is similar to eq.~$(\ref{e9})$ and not to eq.~$(\ref{e7})$. We
thus conclude that eq.~$(\ref{e9})$ and its generalization
eq.~$(\ref{e10})$ are the correct hamiltonians for the interaction of
a scalar test particle with a GW.

\noindent{\bf \S 3. The Free Particle.}

In this section we shall study the interaction of a GW with a free
test particle using the hamiltonian in eq.~$(\ref{e9})$. The solution
of this problem is fairly standard
(see, for example $\cite{Mollow}$). In general terms, due to the
quadrapole nature of the GW the effect of an incident GW is to
produce a squeezed quantum state, which is well known from
quantum optics (see $\cite{ShuCaves}$ for a complete review).
Nevertheless, we shall present a fairly complete analysis of this
system as the analysis of the interaction of the GW with the
two-dimensional harmonic oscillator follows in much the same way.

We first choose the $z$-axis to lie parallel to $\vec k$, the
wave-vector for the GW. Then, due to the transversality
condition eq.~$(\ref{e3})$, $\Gamma^j_{0k}$ has non-zero componants only
in the $x-y$ plane and the particle undergoes free motion along the
$z$-direction. The problem reduces to that of two dimensional motion
with the hamiltonian
\begin{equation}
H= \frac{P_j^2}{2m} + \Gamma^j_{0k}x^jP_k +
\frac{m}{2}\Gamma^j_{0k}\Gamma^j_{0l}x^kx^l\>,
\label{e12}
\end{equation}
where we have used the traceless condition and the indices $j$ and
$k$ now run from $1-2$ only.
Because we are dealing with {\it linearized gravity}, the last term
in eq.~$(\ref{e12})$ dependent on $\Gamma^2$ has little meaning and
we shall set it to zero by hand.

Oftentimes we shall find it more convenient to work with $2\times 2$
matrices representing the two possible polarization states of the GW.
When doing so, the GW is witten as a matrix $h$ with the
explicit componants of $h$ being $h_{jk}$ for $j, k = 1,2$.
Moreover, using the symmetry properties of the GW we can
express
\begin{equation}
h_{jk} = 2f(t) \left(\varepsilon_{\times}
\sigma^1_{jk} + \varepsilon_{+}\sigma^3_{jk}\right) = 2f(t)
\varepsilon_A\sigma^A_{jk}\>,
\label{e13}
\end{equation}
where $\sigma^1$ and $\sigma^3$ are the first and third Pauli
matrices, respectively and the index $A$ runs from $1-3$. $2f(t)$ is
the amplitude of the GW while $\varepsilon_{\times}$ and
$\varepsilon_{+}$ are the polarization states of the wave. They are
in general time dependent and satisfy
\begin{equation}
\varepsilon_{\times}^2+\varepsilon_{+}^2 = 1\>,
\label{e14}
\end{equation}
for all $t$. We shall also assume that the GW incidents the
particle at $t=0$ so that $f(t)=0$ for $t\le0$.

We next define raising and lowering operators
\begin{equation}
x^j \equiv \left({\hbar\over
2m\varpi}\right)^{1/2}\left(a_j+a_j^\dagger\right)\>,\qquad
P_j \equiv -i\left({\hbar m\varpi\over 2}\right)^{1/2}
\left(a_j-a_j^\dagger\right)\>,
\label{e15}
\end{equation}
where $\varpi$ is, as we shall see, related to the uncertainty
in the initial position and momentum of the particle. Then,
\begin{equation}
H = \frac{\hbar\varpi}{2}\left(a_j^\dagger a_j +\frac{1}{2}\right)
    -\frac{\hbar\varpi}{4}\left(a_j^2 + {a_j^\dagger}^2\right)
    -\frac{i\hbar}{4} \dot h_{jk}(t)
    \left(a_ja_k-a_j^\dagger a_k^\dagger\right)\>.
\label{e16}
\end{equation}
Working in the Heisenberg representation, we find that
\begin{eqnarray}
\frac{da_j}{dt} &=& -i\frac{\varpi}{2}(a_j-a^\dagger_j) + \frac{1}{2}\dot
		   h_{jk}a^\dagger_k\>,
\nonumber \\
\frac{da_j^\dagger}{dt} &=& -i\frac{\varpi}{2}(a_j-a^\dagger_j) +
\frac{1}{2}\dot
h_{jk}a_k\>.
\label{e17}
\end{eqnarray}
Next, note that the raising and lowering operators must satisfy the
commutation relations
\begin{equation}
[a_j(t),a_k(t)] = 0\>,\qquad [a_j(t),a^\dagger_k(t)] = \delta_{jk}\>,
\label{e18}
\end{equation}
at equal times. This implies that $a_j(t)$ are related to $a_j(0)$,
the free operators at $t=0$, by the canonical transformation
\begin{eqnarray}
a_j(t) &=& u_{jk} a_k(0) + v_{jk}a^\dagger_k(0)\>,
\nonumber \\
a_j^\dagger(t) &=& a_k^\dagger(0)\bar u_{kj}  + a_k(0)\bar v_{kj}\>,
\label{e19}
\end{eqnarray}
with the bar denoting the complex conjugate.
Eq.~$(\ref{e19})$ is a time-dependent Bogoluibov transformation with
$u$ and $v$ being the generalized Bogoluibov coefficients. They are
$2\times 2$ complex matrices which, due to eq.~$(\ref{e18})$, must
satisfy
\begin{equation}
uv^{t}=u^{t}v\>,\qquad u u^\dagger - v v^\dagger = I
\label{e20}
\end{equation}
written in matrix form. $t$ denotes transpose and $I$ is the identity
matrix. Moreover, they have the boundary conditions $u(0)=I$ and
$v(0) = 0$. Then, from eq.$(\ref{e17})$,
\begin{eqnarray}
\frac{d\>\>}{dt}(u+ v^\dagger) &=& -i\varpi (u-v^\dagger) + \frac{\dot
h}{2}(u+v^\dagger)\>,
\nonumber \\
\frac{d\>\>}{dt}(u-v^\dagger) &=& -\frac{\dot h}{2}(u-v^\dagger)\>.
\label{e21}
\end{eqnarray}

Eq.~$(\ref{e21})$ is difficult to solve for general $h_{jk}$. For
the special cases of linearly and circularly polarized GW, on the
other hand, we can find exact closed form solutions to eq.~$(\ref{e21})$.
In the case of linearly polarized GW, the polarization states
$\varepsilon_{A}$ are constant while $f(t)$ is arbitrary. One can
then directly intergrate eq.~$(\ref{e21})$ to obtain
\begin{eqnarray}
u_l =&{}&\Bigg\{
		\cosh f(t) -i\frac{\varpi}{2}\Bigg(
			\cosh f(t)\int_0^t\cosh(2f(t'))dt'
\nonumber \\
     &{}&		-\sinh f(t)\int_0^t\sinh(2f(t'))dt'
			\Bigg)
	  \Bigg\}I
\nonumber \\
	&{}&-i\frac{\varpi}{2}\Bigg\{
		\sinh f(t)\int_0^t\cosh(2f(t'))dt'
\nonumber \\
	&{}&
			-\cosh f(t)\int_0^t\sinh(2f(t'))dt'
	  \Bigg\}\epsilon_A\sigma^A\>,
\nonumber \\
v_l =&{}&i\frac{\varpi}{2}\Bigg\{
			\cosh f(t)\int_0^t\cosh(2f(t'))dt'
\nonumber \\
	&{}&
			-\sinh f(t)\int_0^t\sinh(2f(t'))dt'
	  \Bigg\}I+
\nonumber \\
	&{}&\Bigg\{
		\sinh f(t) +i\frac{\varpi}{2}\Bigg(
			\sinh f(t)\int_0^t\cosh(2f(t'))dt'
\nonumber \\
	&{}&		-\cosh f(t)\int_0^t\sinh(2f(t'))dt'
			\Bigg)
	  \Bigg\}\epsilon_A\sigma^A\>.
\label{e22}
\end{eqnarray}
For circularly polarized GW, on the other hand, $f(t)=f_0$, a
constant, for $t>0$ while the polarization states now varies with
time
\begin{equation}
\frac{d\epsilon_{+}}{dt} = \Omega
\epsilon_{\times}\>,\qquad\frac{d\epsilon_{\times}}{dt} = - \Omega
\epsilon_{+}\>.
\label{e23}
\end{equation}
The exact solution to eq.~$(\ref{e21})$ in this case is quite
complicated and we shall present here only the small $f_0$ solution
\begin{eqnarray}
u_{cir} =&{}&\left\{
		1-\frac{i\varpi t}{2} + \frac{f^2_0}{2}
		\left(\cosh\Omega t-\cos\Omega t\right)
	     \right\} I
\nonumber \\
	 &{}&-\frac{f_0}{2}\left\{
	     \cosh\Omega t + \cos\Omega t-2 -i\varpi t +
	     i\frac{\varpi}{\Omega}\sin\Omega t
	     \right\}\vec\varepsilon\cdot\vec\sigma
\nonumber \\
	 &{}&+\frac{f_0}{2}\left\{
	 \sinh\Omega t - \sin\Omega t
	 \right\}\frac{\dot{\vec\varepsilon}\cdot\vec\sigma}{\Omega}
\nonumber \\
	 &{}&+\frac{f^2_0}{2}\left\{
	 \sinh\Omega t + \sin\Omega t -2\Omega t
	 \right\}
	 i\frac{\vec\varepsilon\times\dot{\vec\varepsilon}}
	 {\Omega}\cdot\vec\sigma\>,
\nonumber \\
v_{cir} =&{}&\left\{
		\frac{i\varpi t}{2} + \frac{f^2_0}{2}
		\left(\cosh\Omega t+\cos\Omega t-2\right)
	     \right\} I
\nonumber \\
	     &{}&-\frac{f_0}{2}\left\{
	     \cosh\Omega t - \cos\Omega t-2 + i\varpi t -
	     i\frac{\varpi}{\Omega}\sin\Omega t
	     \right\}\vec\varepsilon\cdot\vec\sigma
\nonumber \\
	 &{}&+\frac{f_0}{2}\left\{
	 \sinh\Omega t + \sin\Omega t
	 \right\}\frac{\dot{\vec\varepsilon}\cdot\vec\sigma}{\Omega}
\nonumber \\
	 &{}&-\frac{f^2_0}{2}\left\{
	 \sinh\Omega t - \sin\Omega t
	 \right\}
	 i\frac{\vec\varepsilon\times\dot{\vec\varepsilon}}
	 {\Omega}\cdot\vec\sigma\>,
\label{e24}
\end{eqnarray}
where we have used the vector notation defined in {\bf Appendix A}.
We refer the reader to {\bf Appendix A} for the derivation of the
exact solution.

To complete the quantum mechanical analysis, we note that the unitary
evolution of $a_j(t)$ can be implimented by the following unitary
transformation on $a_j(0)$ and $a_j^\dagger(0)$,
\begin{equation}
a_j(t) = R(t)S(t)a_j(0)S^\dagger(t)R^\dagger(t)\>, \qquad
a_j^\dagger(t) = R(t)S(t)a_j^\dagger(0)S^\dagger(t)R^\dagger(t)\>,
\label{e25}
\end{equation}
where
\begin{equation}
S(t) = \exp\{(\bar\rho_{jk}a_j(0)a_k(0) -
\rho_{jk}a_j^\dagger(0)a_k^\dagger(0))/2\}\>,
\label{e26}
\end{equation}
is the generalization of the squeeze operator of quantum optics, while
\begin{equation}
R(t) = \exp\{-i\epsilon_{jk}a^\dagger_j(0)a_k(0)\}\>,
\label{e27}
\end{equation}
is the generalized rotation operator. $\rho$ and $\tilde\epsilon$ are
$2\times2$ matrices with $\rho$ symmetric: $\rho=\rho^t$. It is then
straightforward to varify that $S$ is indeed
unitary while $R$ is unitary if and only if $\tilde\epsilon
=\tilde\epsilon^\dagger$ is hermitian. The time evolution operator is
then $U(t) = R(t)S(t)$.

Using the canionical commutation relations, we find that $u$ and $v$
must be related to $\rho$ and $\tilde\epsilon$ through
\begin{eqnarray}
u &=& \left(\sum^\infty_{n=0}
	\frac{(\rho\rho^\dagger)^n}{(2n)!}\right) e^{i\tilde\epsilon}\>,
\nonumber \\
v &=& \left(\sum^\infty_{n=0}
	\frac{(\rho\rho^\dagger)^n}{(2n+1)!}\right)\rho
	e^{-i\tilde\epsilon}\>.
\label{e28}
\end{eqnarray}
The constraints eq.~$(\ref{e20})$ requires that $\tilde\epsilon$
also be a real matrix, while $[\rho,\rho^\dagger]=0$.
$\rho$ and $\tilde\epsilon$ are the generalization of the squeeze
parameter and the rotation angle of quantum optics $\cite{ShuCaves}$
with $\rho$ containing within it the squeeze angle also.

The system has now essentially be solved and one need only specify the
initial state of the particle. This involves specifying not only the
initial average position $\langle x^j(0)\rangle$ and momentum
$\langle p^j(0)\rangle$ of the particle, but also its initial
uncertainty in either position and momentum. Since
$\Delta x(0) = \sqrt{\hbar/(2m\varpi)}$ and $\Delta p(0) =
\sqrt{\hbar m\varpi/2}$, doing so determines $\varpi$.

\noindent{\bf \S 4. The Harmonic Oscillator.}

We now consider a GW incident on a two dimensional harmonic
oscillator. Once again for simplicity we shall assume that GW travels
in a direction perpendicular to the oscillating plane of the
oscillator, so that
\begin{equation}
H = \frac{P_j^2}{2m} +
\frac{1}{2}m\omega^2x_j^2 + \Gamma^j_{0k}x^k p_j \>,
\label{e29}
\end{equation}
where $\omega$ is the oscillation frequency of the pendulum, $m$
is its mass and $j,k =1,2$. The oscillator is assumed to have no
equilibrium length in distinct contrast to the system considered by
Weber. We have once again neglected the $\Gamma^2$ term.

We again define raising and lowering operators as in eq.~$(\ref{e15})$, but
with $\omega$ instead of $\varpi$, and obtain
\begin{equation}
H=\hbar\omega(a^\dagger_ja_j+1) - \frac{i\hbar}{2}\dot
h_{jk}\left(a_ja_k-a^\dagger_j a^\dagger_k\right).
\label{e30}
\end{equation}
Note that when $\varepsilon_{+}=1$ and $\varepsilon_{\times}=0$, the
$x$ and $y$ directions decouple and we have the hamiltonian for a one
mode squeeze state (see $\cite{ShuCaves}$) in each direction
seperately. When, on the other hand, $\varepsilon_{+}=0$ and
$\varepsilon_{\times}=1$, the two directions couple and we have
the hamiltonian for a two mode squeeze state.

Proceeding as before, we look for solutions of the
Heisenberg equations of motion with the form eq.~$(\ref{e19})$. We
again obtain a set of matrix differential equations for $u$ and $v$,
but unlike the case of the free test particle, these equations
are no-longer easily solveable even in the case of linearly and
circularly polarized GW.

Let us first consider the case of linearly polarized GW for which
$h_{jk}= 2f(t)\epsilon_A\sigma^A_{jk}\equiv2f(t)\tilde\sigma_{jk}$ and
$\varepsilon_A$ a constant. The equations we now have to solve are
\begin{equation}
\frac{du}{dt} = -iwu + {\dot f}\tilde\sigma v^\dagger\>,\qquad
\frac{dv}{dt} = -iwu + {\dot f}\tilde\sigma u^\dagger\>,
\label{e31}
\end{equation}
along with their corresponding complex conjugate equations. Although
under certain instances these equations can be solved exactly in terms of
Matthieu functions, doing so will not be physically illuminating.
Rather, we note that usually $\vert f(t)\vert\ll1$ and we can solve
eq.~$(\ref{e31})$ perturbatively about its  $f(t)=0$ solution
\begin{eqnarray}
u_l\approx&{}& e^{-iwt}\Bigg(1+\frac{f^2(t)}{2} +
4\omega^2\int^t_0f(t')\int^{t'}_0f(t'')	e^{2i\omega(t'-t'')}dt'dt''
\nonumber \\
 	&{}& +2i\omega f(t)\int^t_0f(t')e^{2i\omega(t-t')}dt
	-2i\omega\int^t_0f^2(t')dt'\Bigg)I\>,
\nonumber \\
v_l\approx&{}& e^{iwt}\left(f(t) -
	2i\omega\int^t_0f(t')e^{-2i\omega(t-t')}dt'\right)\tilde\sigma\>.
\label{e32}
\end{eqnarray}

As we have mentioned, the hamiltonian for the interaction of a GW with
a harmonic oscillator is very similar to that of the hamiltonian for
squeezed states in quantum optics. Indeed, if we solve
eq.~$(\ref{e28})$ in this case, we find that $\rho =
re^{2i\phi}\tilde\sigma$ and $\tilde\epsilon = \epsilon I$ where
\begin{eqnarray}
\tan(\epsilon-wt) = &{}&
		2\omega f(t)\int^t_0 f(t')\cos(2\omega(t-t'))dt'
		-2\omega\int^t_0f^2(t')dt'
\nonumber \\
		&{}&
		+4\omega\int^t_0f(t')\int^{t'}_0
		f(t'')\sin(2\omega(t'-t''))dt'dt''\>,
\nonumber \\
\tan(\epsilon-2\phi+wt) = &{}&
		\frac{2\omega\int^t_0f(t')\sin(2\omega(t-t'))dt'}
		{f(t) - 2\omega\int^t_0f(t')\sin(2\omega(t-t'))dt'}\>.
\label{e33}
\end{eqnarray}
There is, however, some ambiguity in the solution for $r$. If one uses
the equation for $u$, one finds that
\begin{eqnarray}
r^2 = &{}&f^2(t) - 4\omega f(t)\int^t_0 f(t')\sin(2\omega(t-t'))dt'
\nonumber \\
      &{}&+ 8\omega^2\int^t_0f(t')\int^{t'}_0f(t')\cos(2w(t'-t''))dt'dt''\>,
\label{e34}
\end{eqnarray}
while if we use the equation for $v$,
\begin{equation}
r^2 = f^2(t) - 4\omega f(t)\int^t_0 f(t')\sin(2\omega(t-t'))dt'
	+ 8\omega^2\left(\int^t_0f(t')\cos(2w(t-t'))dt'\right)^2\>.
\label{e35}
\end{equation}
This ambuigity araises from our perturbation scheme and the
desire to polar decompose $u$ and $v$ into a modulus and phase. The
differences between the two expressions only occur in
the last term, however, and are very small for $wt\ll1$. Any ambiguity
disappears if we use eq.~$(\ref{e28})$ in eq.~$(\ref{e31})$ to obtain
differential equations for $\rho$ and $\tilde\epsilon$ and then
preform the perturbative analysis on $\rho$ and $\tilde\epsilon$
directly. For our purposes, however, the perturbative solutions
eqs.~$(\ref{e33})$ and $(\ref{e34})$ are sufficient.

The parameters $r$, $\epsilon$ and $\phi$ are usually called the
squeeze parameter, the rotation angle and the squeeze angle,
respectively. There are two properties of $r$ and $\phi$ that are of
interest. First, from eq.~$(\ref{e32})$ we expect a resonance to
occur when the frequency of the GW is equal to $2\omega$. This can be
seen explicitly by taking $f(t) = f_0\sin\Omega t$ for a monochromatic
GW and performing the integral in eq.~$(\ref{e32})$ explicitly.
Physically, this is due to the quadrapole nature of the GW, and its
quadratic coupling to the harmonic oscillator. At this order of the
perturbative expansion we get frequency doubling, and thus a
resonance at $\Omega=2\omega$ for $u$ and $v$. If we were to then go
to the next perturbative order, we would get frequency quadrupoling
and a resonance at $\Omega=4\omega$ and so on.

Second, notice that to this order {\it $\phi$ is independent of the
amplitude of the GW}. Consequently, a very weak GW can still produce
a large response in the squeeze angle. Indeed, for the monochromatic
wave $f(t)=f_0\sin\Omega t$,
\begin{equation}
\tan(\epsilon-2\phi+wt) =-2\omega
		\frac{\omega\sin\Omega t-\Omega\sin2\omega t}
		{(\Omega^2-2\omega^2)\sin\Omega t
		 -2\omega\Omega\sin2\omega t}\>,
\label{e36}
\end{equation}
when $\Omega\ne2\omega$. Note that the denominator of eq.~$(\ref{e36})$ may
vanish for certain $t$. (This will, in fact, almost certainly happen when
if the incident GW is on resonance $\Omega=2\omega$.) Consequently, we
can expect very large variations in the squeeze angle $\phi$ no
matter how weak the incident GW is. Since, however, the squeeze
parameter $r\sim f(t)$, this large phase shift will still be
difficult to see. Note also that this does {\it not\/} happen for
$\epsilon$, which depends on $f^2$ and to lowest order is unaffected
by the GW.

Finally, let us consider a circularly polarized GW for which
\begin{equation}
\frac{du}{dt} = -iwu + f_0{\dot \varepsilon_A}\sigma^A v^\dagger\>,\qquad
\frac{dv}{dt} = -iwu + f_0{\dot \varepsilon_A}\sigma^A u^\dagger\>.
\label{e37}
\end{equation}
Although these equations can be solved exactly using the techniques
described in {\bf Appendix A}, the final result will not useful as it
involves the solution of a fourth order polynomial equation. We shall
instead once again solve eq.~$(\ref{e37})$ perturbatively and obtain
\begin{eqnarray}
u_{cir} = &{}&e^{-i\omega t}
	  \Bigg\{
	  	1-2i\omega f_0
\int^t_0\vec\varepsilon(t)\cdot\vec\varepsilon(t') e^{2i\omega(t-t')}dt'
\nonumber \\
	&{}&
			+4\omega^2\int^t_0\int^{t'}_0\vec\varepsilon(t')
			\vec\varepsilon(t'')e^{2i\omega(t'-t'')}dt'dt''
	  \Bigg\}I
\nonumber \\
	&{}&
	+
	e^{-i\omega t}f_0^2\left\{2i\frac{\omega}{\Omega}
	\int^t_0 i\frac{\vec\varepsilon(t)\times\dot{\vec\varepsilon(t)}}
	{\Omega}\cdot\vec\varepsilon(t')e^{2i\omega(t-t')}dt'\right\}
	\frac{\dot{\vec\varepsilon}\cdot\vec\sigma}{\Omega}
\nonumber \\
	&{}&
	  +e^{-i\omega t}f_0^2
	  \Bigg\{
	  -\Omega t +
	  -2i\frac{\omega}{\Omega}\int_0^t\dot{\vec\varepsilon}(t)
	  \cdot\vec\varepsilon(t')e^{2i\omega(t-t')}dt'
\nonumber \\
	&{}&
	  +4i\frac{\omega^2}{\Omega^2}\int^t_0\int^{t'}_0
	  \bigg(\vec\varepsilon(t)\cdot\vec\varepsilon(t')
	  \dot{\vec\varepsilon(t)}\cdot\vec\varepsilon(t'')
\nonumber \\
	&{}&
	  -
	  \vec\varepsilon(t)\cdot\vec\varepsilon(t'')
	  \dot{\vec\varepsilon(t)}\cdot\vec\varepsilon(t')\bigg)
	  e^{2i\omega(t'-t'')}dt'dt''
	  \Bigg\}i\frac{\vec\varepsilon\times
	  \dot{\vec\varepsilon}}{\Omega}\cdot\vec \sigma
\nonumber \\
v_{cir} = &{}& f_0e^{i\omega t}\left(
		1+2i\omega\int^t_0\vec\varepsilon(t)
		\vec\varepsilon(t')e^{-2i\omega(t-t')}dt'
		\right)\vec\epsilon\cdot\vec\sigma
\nonumber \\
	&{}&
		+
		2if_0\frac{\omega}{\Omega}e^{i\omega t}\left(
		\int^t_0\dot{\vec\varepsilon(t)}\cdot\vec\varepsilon(t')
		e^{-2i\omega(t-t')}dt'
		\right)\frac{\dot{\vec\epsilon}\cdot\vec\sigma}{\Omega}\>.
\label{e38}
\end{eqnarray}
We once again expect a resonance at $\Omega=2\omega$.
The corresponding $\rho$ and $\tilde\epsilon$ can also be found in
this case.

\noindent{\bf \S 5. The Hydrogen Atom.}

As we shall be using time dependent perturbation theory to analyze
the interaction of GW with the hydrogen atom, in this section we
shall keep to generalities. An exposition of time dependent
perturbation theory can be found in any quantum mechanics textbook
(see for example $\cite{Merz}$). From eq.~$(\ref{e10})$ the relevant
hamiltonian is
\begin{eqnarray}
H = &{}&\frac{p_j^2}{2\mu} +V-\frac{e}{\mu c}A_j p_j +\frac{e^2}{2\mu
    c^2}A_j^2
\nonumber \\
	&{}&
    -\frac{e}{c}\Gamma^j_{0k}A_jx^k +
    \Gamma^j_{0k}x^kp_j +\frac{m}{2}\Gamma^j_{0k}\Gamma^j_{0l}x^kx^l
\label{e39}
\end{eqnarray}
where $\mu$ is the reduced mass of the hydrogen atom, $A_j$ is the
vector potential and $V$ is the usual Coulomb potential for the
hydrogen atom. Although the vector potential will be treated as a field
operator, the energy density for the em-field has not been
included in eq.~$(\ref{e39})$. We have moreover neglected the spin
of the electron, as we have not yet considered the effect of the GW
on the spin of a particle. We also note that even at atomic energies,
the wave-length of the GW will still be much larger than the size of
the hydrogen atom and the long wave-length approximation is still
valid.

The first four terms in eq.~$(\ref{e39})$ are what one usually obtains
for a minimally coupled hydrogen atom. The additional three terms
comes from the interaction of the GW with the atom and we
shall treat these terms  perturbatively. As usual, since the last
term involves $\Gamma^2$ we shall not consider its affects.

Consider first the interaction term
\begin{equation}
H^{(1)}_{int} = \frac{1}{2}\dot h_{jk}x^kp_j\>,
\label{e40}
\end{equation}
which couples the GW directly to the electron and
can cause either the absorption or emission of a ``graviton''.
{}From time-dependent perturbation theory the transition probability
between an initial state $\vert init\rangle$ and a final state $\vert
fin\rangle$ is
\begin{equation}
\vert c(+\infty)\vert^2 = \frac{\mu^2\pi^2}{4\hbar^2}\omega_{if}^4
		\vert \hat h_{jk}(\omega_{if})\vert^2
		\vert\langle fin\vert \left(x_jx_k-\delta_{jk}x^2/3\right)
		\vert init\rangle\vert^2
\label{e41}
\end{equation}
where $\hat h_{jk}(\omega_{if})$ is the fourier transform of $h_{jk}(t)$ and
$\hbar\omega_{if}$ is the energy difference between the initial and
final states. The quadrapole nature of the interaction can now be seen
explicitly. From the Wigner-Eckhart theorem the allowed
transitions are: $\Delta m = \pm 2$, $\Delta l = \pm 2$ where $m$ and
$l$ are the angular momentum quantum numbers.

Using eq.~$(\ref{e41})$ and the fact that the intensity of the GW is
$I = h_{jk}^2)\omega^2 c^3/32\pi G$, find that the
cross-section for this interaction has the form
\begin{equation}
\sigma_{cross}(\omega) = J_{if} l_p^2\>,
\label{e42}
\end{equation}
where $J_{if}$ is a numerical factor depending on the initial and final
states and $l_p^2 = G\hbar/c^3$ is the Planck length. (The reader
should not confuse this $\sigma_{cross}$ with the Pauli matrices.)
The cross-section is thus proportional to $l_p^2=10^{-66}$ cm$^2$,
and is extremely small no matter what value $J_{if}$ takes. Combined
with the expectation that the intensity of GW's in the universe
which are at atomic energies are very low, such a small cross-section means
that it will essentially be impossible to see a GW induced transition in a
hydrogen atom. Nevertheless, it is interesting to see the appearance of
the Planck scale even at this simplestic level. Notice
also the absence of knowledge of the em nature of the system, such as
the charge of either the electron or the nucleus, in
eq.~$(\ref{e42})$. This once again underscores the fact that
$H_{int}^{(1)}$ is due to the direct interaction of gravity with matter.

Let us next consider the interaction term
\begin{equation}
H^{(2)}_{int} = \frac{e}{2c}\dot h_{jk}A_jx^k\>.
\label{e43}
\end{equation}
In may ways this is much more interesting as it couples
the GW with both the em-field and matter. In particular, this is a
{\it dipole\/} coupling in contrast to the
quadrapole coupling of $H^{(1)}_{int}$. The absorption of a
graviton can thus cause the emission of a photon along with the excitation
of the hydrogen atom to an excited state. Conversely, the emission of
a photon can also cause the emission of a graviton. Theoretically,
the passage of a GW should shorten the lifetime of the any excited
state of the atom. Indeed, if we combined eq.~$(\ref{e43})$ with the
usual dipole interaction term in eq.~$(\ref{e39})$, we find that the
cross-section for emission or absorption of a photon in the dipole
approximation will be slightly larger by a factor $1+const.
h_{jk}^2(0)$ where the constant depends on the polarization state of
the GW. Since, however, $h_{jk}$ is extremely small, this shortening
cannot be seen experimentally.

Using Fermi's Golden Rule, we can also calculate the transition
cross-section for the emission or absorption of a single photon due
to $H^{(2)}_{int}$,
\begin{equation}
\sigma_{cross}(\omega) = J_{if}' \alpha \left(\omega_{if}
a_{bohr}/c\right)^2 l_p^2\>,
\label{e44}
\end{equation}
where once again $J_{if}'$ is a numerical factor depending on the
initial and final states of the hydrogen atom, $\hbar\omega_{if}$ is
the energy of the photon, $a_{bohr}$ is the Bohr radius and $\alpha$
is the fine structure constant. For typical atomic transitions,
$\sigma_{cross}\sim 10^{-5}l_p^2$ which is even smaller than that of
the direct interaction $H^{(1)}_{int}$ since the transition is
mediated by the emission of a photon.

\noindent{\bf \S 7. Concluding Remarks.}

To conclude, we have quantized the classical equations of motion for a
scalar, spin-$0$ interacting with a classical GW in the long
wavelength limit when the velocities of the particle is small. We
find that due to the quadrapole nature of the interaction the effect
of the GW is to produce a squeezed quantum state
\footnote{Curiously, the creation of primordial GW from the de
Sitter vacuum is also due to quantum mechanical squeezing
$\cite{Gris2}$.}.
We have then used this formalism to calculate the effects of the GW
on the free particle, the harmonic oscillator and the hydrogen atom.

The hamiltonian $H$ that we have obtained is quite peculiar, however.
On the one hand it has the form one
would expect if the analogy between the GW and the em-wave is to hold
at this level. Like the em-wave, the GW carries momentum and energy.
And like the em-wave, since the classical equations of motion depends
on a physical observable (the field strength for the em-wave, the
curvature tensor for the GW), the hamiltonian must depend on their
respective connections. A coupling of the form
$(P_j+m\Gamma^j_{0k}x^k)^2$ would thus seem quite natural. Moreover, like
the vector potential for the em-wave it means that the connection
$\Gamma^j_{0k}$ now takes on an independent meaning at the quantum
mechanical level. On the other hand, this coupling resembles minimal
coupling for the gravitational field. As we are dealing with a scalar
particle, from general principles alone we would not have expected a coupling
of this form. We would instead have expected a hamiltonian
of the form given by DeWitt $\cite{DeWitt}$. One cannot, however,
obtain $H$ directly from DeWitt's hamiltonian by taking the
linearized gravity limit.

DeWitt in $\cite{DeWitt}$ considered the quantum mechanics of a
particle constrainted to move on a curved space(time). Then if
$g_{ij}$ is the metric {\it on the three dimensional hypersurface\/}
he obtained a hamiltonian of the form
\begin{equation}
H_{DW} = \frac{1}{2m}g^{jk}\left(-i\hbar\frac{\partial\>}{\partial x^j}
-\frac{i\hbar}{4}\frac{\partial\ln g}{\partial
x^j}\right)\left(-i\hbar\frac{\partial\>}{\partial
x^k}-\frac{i\hbar}{4}\frac{\partial\ln g}{\partial x^k}\right)
\label{e46}
\end{equation}
for a non-relativistic particle where $g$ is the
determinant of $g_{jk}$.
\footnote{DeWitt considered motion on an
$n$-dimensional space. We need not be so general.} Let us now try to
apply this hamiltonian to the case of GW. In this case $g_{jk} =
\delta_{jk}+h_{jk}$ and due to the traceless condition on $h_{jk}$,
$g=1$ up to first order in $h_{jk}$. Consequently,
\begin{equation}
H_{DW} \approx -\frac{\hbar^2}{2}\left(\frac{\partial\>}{\partial
x^j}\right)^2 -\frac{\hbar^2}{2}h^{jk}\frac{\partial\>}{\partial
x^j}\frac{\partial\>}{\partial x^k}\>.
\label{e47}
\end{equation}
This hamiltonian is
very different from only the one we have derived, but also from
the one used by Weber. Moreover, from the transformation law for
general coordinate transformations (eq.~$(5.42)$ of $\cite{DeWitt}$),
$H_{DW}$ cannot be mapped to $H$ through a judicious coordinate
transformation.

We should also mention that in DeWitt's derivation of
$H_{DW}$ he actually obtained $p_j = -i\hbar(\partial_j+\partial_j
\ln g/4) - \partial_j\Phi$ where $\Phi$ is an arbitrary real function
of $x_j$ and $t$ (eq.~$(5.24)$ of $\cite{DeWitt}$). $\Phi$ was then
set to zero by arguing that it can always be removed by a {\it local\/}
unitary phase transformation on the wavefunction. One finds precisely
a term of this form in $H$. This, however, still does not explain the
presence of the second term in $H_{DW}$ which is absent in $H$.

We believe that the differences between $H_{DW}$ and $H$
to be more fundamental than a simple gauge transformation of the
wavefunction, however. DeWitt's derivation of $H_{DW}$ was based on a
natural generalization of quantum mechanics on flat spacetime to curved
spacetimes. Our hamiltonian, on the other hand, is based on the
classical geodesic {\it deviation\/} equation. Physically, this is
because the passage of the GW will affect not
only the test particle, but also the observer (measuring apparatus,
laboratory setup). Since the observer is
also part of the universe, it cannot be isolated from any
gravitational effects and also responses to the GW. Consequently, one
cannot measure (observe) the {\it absolute\/} motion of the test
particle, but rather the {\it relative\/} motion between the observer and
the test particle. This motion is governed by the geodesic deviation
equation. In DeWitt's formalism it is not certain where the
observer is. Since he deals with only a single particle, we suspect
that the observer has tacitly been ``removed'' from the spacetime,
something one can do in Newtonian mechanics, but cannot do in general
relativity. As the results of this paper is only valid within the
long wave-length limit and for a fixed choice of gauge we have not
yet been able to show this explicitly.

Further study of the interaction of GW with matter is clearly needed
to clarify these and other issues. In particular, a version of the
hamiltonian which is valid outside of the long wave-length
approximation is needed and the effect of GW on particles with spin
should also be incorporated $\cite{ADS}$. Once this is accomplished,
a thorough comparison of our formalism and DeWitt's results can be
made.

We end this paper by briefly mentioning the differences between the
hamiltonian used by Weber and the one we derived. As the
work by Weber was mostly concerned with the properties of the
interaction of GW with gravitational wave detectors, he had at
their disposal a macroscopic length $l$. For the beam detector this
is the length of one arm of the interferometer, while for the bar
detector this is the equilibrium length of the spring for a harmonic
oscillator. In both cases, the hamiltonian that was used has the form
\begin{equation}
H_{Web} = \frac{p^2}{2m} + \frac{m}{2}\omega^2x^2 + mR^1_{0,10}xl\>,
\end{equation}
along, say, the $x$-direction. (The harmonic oscillator term is absent of
the beam detector). Importantly, $x$ is measured from its {\it
equilibrium\/} position. If we replace $x\to x+l$ in eq.~$(\ref{e7})$ and
expand out the quadratic term, we obtain $H_{Web}$ up to a term
proportional to $x^2$. This was dropped in comparison to $xl$ since
the response to the GW is expected to be quite small classically.
While this is perfectly valid in classical dynamics, since $x$ is now
an operator doing so is somewhat questionable at a quantum mechanical level.
We have, however, re-preformed the analysis of both the beam and the
bar detectors using the full hamiltonian eq.~$(\ref{e9})$ in a
quantum mechanics framework and found very little differences in the
results obtained from using $H_{Web}$ verses $H$. Consequently, as
long as the linearization of eq.~$(\ref{e9})$ is valid, and as long as
the test particle is not charged, the hamiltonian used by Weber and
others gives a good description of the interaction of a GW with
matter.

\begin{appendix}
{}
We first define $\xi = u+v^\dagger$ and $\chi = u - v^\dagger$ so
that eq.~$(\ref{e21})$ becomes
\begin{equation}
\frac{d\xi}{dt} = -i\varpi\chi + f_0\dot\varepsilon_A\sigma^A\xi\>,
\qquad
\frac{d\chi}{dt} = -f_0\dot\varepsilon_A\sigma^A\chi\>,
\label{a1}
\end{equation}
for circularly polarized GW.
Next, we note that any $2\times2$ complex matrix $M$ can be written has
$M = \theta_0 I + \theta_A\sigma^A$ where $\theta_0$ and $\theta_A$
are complex numbers. We next consider $\vec\theta = (\theta_1,
\theta_2,\theta_3)$ as being a vector in a three dimensional complex
space. Clearly the polarization states of the GW can also be represented
as a vector $\vec\varepsilon$ in this space. Moreover,
$\vec\varepsilon$, $\dot{\vec\varepsilon}$ and
$\vec\varepsilon\times\dot{\vec\varepsilon}$ are mutually orthogonal
and thus form a natural coordinate system for this space.
Consequently, we look for solutions of (\ref{a1}) with the form
\begin{equation}
\chi = A I + B\vec\varepsilon\cdot\vec\sigma +
       C\frac{\dot{\vec\varepsilon}\cdot\vec\sigma}{\Omega} +
       Di\frac{\vec\varepsilon\times
       \dot{\vec\varepsilon}}{\Omega}\cdot\vec\sigma\>,
\end{equation}
where $A$, $B$, $C$, $D$ can be complex functions. Then
\begin{eqnarray}
\frac{dA}{dt} + f_0\Omega C &=& 0\>,
\nonumber \\
\frac{dB}{dt} - \Omega C-f_0\Omega D &=& 0\>,
\nonumber \\
\frac{dC}{dt} + \Omega B + f_0\Omega A &=& 0\>,
\nonumber \\
\frac{dD}{dt} - f_0\Omega B &=& 0\>.
\end{eqnarray}
Next, using the boundary condition $\chi(0)=u(0)=I$, we find that
\begin{eqnarray}
A(x) &=& \frac{\nu_{+}^2\cos\nu_{-}x - \nu_{-}^2\cos\nu_{+}x +
f_0^2(\cos\nu_{-}x -\cos\nu_{+}x)}{\nu_{+}^2-\nu_{-}^2}\>,
\nonumber \\
B(x) &=&
-\frac{1}{f_0}\frac{(\nu_{+}^2+f_0^2)(\nu_{-}^2+f_0^2)}{\nu_{+}^2-\nu_{-}^2}
\left(\cos\nu_{-}x - \cos\nu_{+}x\right)\>,
\nonumber \\
C(x) &=& \frac{1}{f_0}\frac{\nu_{+}\nu_{-}(\nu_{+}\sin\nu_{-}x -
\nu_{-}\sin\nu_{+}x) + f_0^2(\nu_{-}\sin\nu_{-}x
-\nu_{+}\sin\nu_{+}x)}{\nu_{+}^2-\nu_{-}^2}\>,
\nonumber \\
D(x) &=& -\frac{(\nu_{+}^2+f_0^2)(\nu_{-}^2+f_0^2)}{\nu_{+}^2-\nu_{-}^2}
\left(\frac{\sin\nu_{-}x}{\nu_{-}} -\frac{\sin\nu_{+}x}{\nu_{+}}\right)\>,
\end{eqnarray}
where $x=\Omega t$ and
\begin{equation}
\nu_{\pm}^2\equiv\frac{1}{2}\left(1-2f_0^2\pm\sqrt{1-4f_0^2}\right).
\end{equation}

The solution for $\xi$ follows in exactly the same way. We once again
take
\begin{equation}
\xi =  E I + F\vec\varepsilon\cdot\vec\sigma +
       G\frac{\dot{\vec\varepsilon}\cdot\vec\sigma}{\Omega} +
       Hi\frac{\vec\varepsilon\times
       \dot{\vec\varepsilon}}{\Omega}\cdot\vec\sigma\>,
\end{equation}
and now find that
\begin{eqnarray}
E(x) &=& \frac{\kappa_{+}^2\cosh\kappa_{-}x -
\kappa_{-}^2\cosh\kappa_{+}x + f_0^2(\cosh\kappa_{+}x
-\cosh\kappa_{-}x)}{\kappa_{+}^2-\kappa_{-}^2} +
\frac{i\varpi C}{f_0\Omega} + \frac{i\varpi D}{f_0^2\Omega} \>,
\nonumber \\
F(x) &=&
\frac{1}{f_0}\frac{(\kappa_{+}^2-f_0^2)(\kappa_{-}^2-f_0^2)}
{\kappa_{+}^2-\kappa_{-}^2} \left(\cosh\kappa_{+}x -
\cosh\kappa_{-}x\right) +\frac{i\varpi D}{f_0\Omega} \>,
\nonumber \\
G(x) &=& \frac{1}{f_0}\frac{\kappa_{+}\kappa_{-}(\kappa_{+}\sinh\kappa_{-}x -
\kappa_{-}\sinh\kappa_{+}x) + f_0^2(\kappa_{+}\sinh\kappa_{+}x
-\kappa_{-}\sinh\kappa_{-}x)}{\kappa_{+}^2-\kappa_{-}^2}\>,
\nonumber \\
H(x) &=&
-\frac{(\kappa_{+}^2-f_0^2)(\kappa_{-}^2-f_0^2)}{\kappa_{+}^2-\kappa_{-}^2}
\left(\frac{\sinh\kappa_{+}x}{\kappa_{+}}
-\frac{\sinh\kappa_{-}x}{\kappa_{-}}\right)\>,
\end{eqnarray}
with
\begin{equation}
\kappa_{\pm}^2\equiv\frac{1}{2}\left(1+2f_0^2\pm\sqrt{1+4f_0^2}\right).
\end{equation}
Notice that for $f_0\ll 2$, both $\nu_{\pm}$ and $\kappa_{\pm}$ are
real. Eq.~$(\ref{e24})$ now follows if we take the $f_0\to0$ limit
and keep terms up to $O(f^2_0)$.
\end{appendix}

\begin{center}
{\bf Acknowledgements}
\end{center}

This work was supported in part by the R.O.C. NSC Grant No.
NSC83-0208-M001-69.

\end{document}